\documentclass[a4paper,11pt]{article}
\usepackage{pos}

\newcommand{\La}{{\Lambda}}
\newcommand{\Si}{{\Sigma}}
\newcommand{\ccc}{\cdot\cdot\cdot}
\newcommand{\be}{\begin{eqnarray}}
\newcommand{\ee}{\end{eqnarray}}

\title{Strangeness $\bf{S=-3}$ and $\bf{-4}$ baryon-baryon interactions in chiral effective
field theory}
\ShortTitle{Strangeness ${S=-3}$ and ${-4}$ baryon-baryon interactions in chiral EFT}

\author*[a]{J. Haidenbauer}
\author[b,a,c]{U.-G. Mei{\ss}ner}

\affiliation[a]{Institute for Advanced Simulation, Forschungszentrum J\"ulich GmbH,\\ 
D-52428 J\"ulich, Germany}

\affiliation[b]{Helmholtz-Institut f\"ur Strahlen- und Kernphysik and 
Bethe Center for Theoretical Physics,\\ University of Bonn, 
D-53115 Bonn, Germany}

\affiliation[c]{Tbilisi State  University,  0186 Tbilisi, Georgia}

\emailAdd{j.haidenbauer@fz-juelich.de}
\emailAdd{meissner@hiskp.uni-bonn.de}

\abstract{Studies of the baryon-baryon interaction involving hyperons within
chiral effective field theory, so far performed up to next-to-leading 
order (NLO) in the chiral expansion, 
have shown that for the strangeness $S=-1$ ($\Lambda N$, $\Sigma N$)
and $S=-2$ ($\Lambda \Lambda$, $\Xi N$) sectors a consistent and satisfactory 
description of the available scattering data and experimental constraints can 
be achieved based on the assumption of broken SU(3) flavor symmetry.
We explore the possible extension of this approach at the NLO level 
to strangeness $S=-3$ and $S=-4$ baryon-baryon systems where empirical 
information is rather scarce. Specifically we address the question how 
measurements of two-particle correlation functions in heavy-ion collisions 
and/or in high-energetic proton-proton collisions can help to constrain 
the interaction in channels like $\Xi\Lambda$ or $\Xi\Xi$.
}

\FullConference{%
  The 10th International Workshop on Chiral Dynamics - CD2021\\
  15-19 November 2021\\
  Online
  }

\begin{document}
\maketitle

\section{Introduction}

Chiral effective field theory ($\chi$EFT) is a powerful tool for calculating 
the dynamics of baryons in low-energy hadronic processes \cite{Epelbaum:2009}. 
This approach incorporates the same symmetries and symmetry breaking 
patterns at low energies as QCD, but it builds on the relevant physical 
degrees of freedom, namely hadrons, instead of quarks and gluons.
Based on an expansion in small external momenta in combination with an appropriate power 
counting, the results can be improved systematically by going to higher orders,
and at the same time theoretical uncertainties can be estimated.
Furthermore, two- and three-baryon forces can be constructed in a consistent way. 
The approach leads to an interaction potential which can be readily employed in
standard two- and few-body calculations. It consists of contributions 
from an increasing number of pseudoscalar-meson exchanges, determined 
by the underlying chiral symmetry,  
while the short-distance dynamics remains unresolved and is encoded in 
contact terms whose strengths are parameterized by a priori unknown 
low-energy constants (LECs).
For the nucleon-nucleon ($NN$) system which is experimentally very well 
studied, $NN$ potentials have been derived up to high order in the chiral 
expansion which describe the scattering data with remarkably high precision
\cite{Reinert:2017,Entem:2017}.  

The J\"ulich-Bonn-Munich Collaboration has applied $\chi$EFT  
to investigate baryon-baryon (BB) interactions involving hyperons.
These studies, performed so far up to next-to-leading order (NLO) in the chiral
expansion, have shown that for the strangeness $S=-1$ ($\Lambda N$, $\Sigma N$) 
\cite{YN2013,YN2019} and $S=-2$ ($\Lambda \Lambda$, $\Xi N$) \cite{YY2015,YY2019} 
sectors a consistent and satisfactory description of the available scattering 
data and other experimental constraints can be achieved within 
the assumption of (broken) SU(3) flavor symmetry. In addition, applications of 
the resulting potentials in bound-state calculations for light hypernuclei led to 
results close to the empirical values \cite{YN2019,Le:2020,Le:2021LL}.   

In this work we explore the possible extension of this approach 
to strangeness $S=-3$ and $S=-4$ BB systems where there is 
practically no experimental information. 
Until now solely lattice QCD simulations \cite{Ishii:2018,Doi:2018,NPLQCD:2021} 
have provided a glimpse on such BB interactions.
A full extension of chiral EFT to the $S=-3$ and $-4$ has been considered only 
at leading order (LO) \cite{XX2010,Liu:2020}. At this order all the ocurring LECs
can be inferred from studies of the $\Lambda N$ and $\Sigma N$ systems by exploiting 
the underlying SU(3) flavor symmetry. However, the resulting LO potentials 
turned out to be strongly attractive, quite at odds with those lattice QCD 
predictions, which very likely signals a non-negligible SU(3) symmetry breaking. 
Indeed, already at NLO terms involving external fields arise which lead to genuine 
SU(3) symmetry breaking contact terms~\cite{Petschauer:2013}. Those have
been taken into account in a first attempt for an extension to $S=-3,-4$ 
at the NLO level~\cite{XX2015}, which exploited the feature that for the $^1S_0$ 
partial wave and specific channels the number of SU(3) symmetry breaking terms is small.
Specifically, the interactions in the $NN$, $\Si N$, $\Si \Si$, $\Xi \Si$, 
and $\Xi\Xi$ channel with maximal isospin are identical within strict SU(3) symmetry, 
and they involve only two additional independent SU(3) symmetry breaking LECs. 
The situation for the $^3S_1$ state is much more complicated because it involves 
different combinations of SU(3) symmetry preserving and breaking contact terms
in basically all reaction channels. Thus, without experimental constraints in 
the $S=-3$ and $-4$ systems fixing the corresponding LECs is practically 
impossible. 

Very recently a new doorway to BB interactions involving strangeness has 
been opened, in form of two-particle momentum correlation functions \cite{Cho:2017} 
that can be measured in heavy-ion collisions and/or in high energetic 
proton-proton collisions. Systems like $\La p$, $\La\La$, or 
$\Xi^-p$ 
\cite{Adams:2006,Adamczewski:2016,ALICE:2021L,STAR:2015,ALICE:2019S,ALICE:2019,ALICE:2020} 
have been already successfully investigated over the last decade. However, 
now also correlation-function measurements for channels like $\Xi\Lambda$ 
\cite{ALICE:2021} or $\Xi\Xi$ \cite{STAR:2021} are on their way. 
In view of this development, in the present work we study how that 
information could help to constrain the interaction in the $S=-3$ and $-4$ 
sectors. For that purpose, we consider extensions of BB interactions 
whose contact terms have been fixed in line with the available constraints 
on the $\La\La$ and $\Xi N$ systems and 
explore the predictions for such correlation functions. 

\section{Baryon-baryon interaction in SU(3) chiral effective field theory} 

The treatment of the BB interaction within SU(3) $\chi$EFT is described in detail 
in Refs.~\cite{YN2013,YY2015,Polinder:2006,Polinder:2007,Petschauer:2020}. 
Specifically, the formalism for deriving the potential up to
NLO for strangeness $S=-1$ ($\La N$, $\Si N$) is provided in \cite{YN2013}
whereas the extension to $S=-2$ ($\La\La$, $\Xi N$, $\La\Si$, $\Si\Si$) 
can be found in \cite{YY2015}. The application to $S=-3$ and $-4$ is only documented 
for the LO case \cite{XX2010}, but is straightforward for NLO. Indeed, the structure
of the (irreducible) two-meson contributions is identical to the one given in 
our work on the $\La N$ and $\Si N$ interactions \cite{YN2013}. Only the 
coupling constants at the meson-baryon-baryon vertices, given by the standard 
SU(3) relations, change since the nucleon is replaced by the $\Xi$.   
The structure of the contact terms is similar too, at least for the SU(3) 
symmetric part, except that the role of the $\{10\}$ and $\{10^*\}$ irreps
of SU(3) is interchanged \cite{XX2010}. This concerns the $^3S_1$ 
state and, of course, all other spin-space symmetric partial waves. With the SU(3) 
symmetry breaking contact terms included, the relations become
more complex \cite{Petschauer:2013}. To be concrete, the contributions from 
the contact terms up to NLO are of the general form
\begin{eqnarray}
V(^1S_0) 
&=& {\tilde{C}}_{^1S_0} + {C}_{^1S_0} ({p}^2+{p}'^2) + {C}_{^1S_0}^{\chi} (m^2_{K}-m^2_{\pi}), \nonumber \\
V(^3S_1) 
&=& {\tilde{C}}_{^3S_1} + {C}_{^3S_1} ({p}^2+{p}'^2) + {C}_{^3S_1}^{\chi} (m^2_{K}-m^2_{\pi}), \nonumber \\
V(\alpha) &=& {C}_{\alpha}\, {p}\, {p}' ~,\quad  \alpha \, \in \, \{^3P_1, \ ^1P_1, \ ^3P_1, 
\ ^1P_1 -\, ^3P_1, \ ^3P_2\}, \nonumber\\
V(^3D_1 \leftrightarrow\, ^3S_1) &=& {C}_{^3S_1 -\, ^3D_1}\, {p'}^2, \  {C}_{^3S_1 -\, ^3D_1}\, {p}^2, \ 
\label{Eq:SU31} 
\end{eqnarray}
with $p$ and $p'$ the center-of-mass momenta of the initial and final BB state. 
$\tilde C_\alpha$ and $C_\alpha$ generically denote LECs that correspond to SU(3) 
symmetric contact terms and, for each BB channel and partial wave, are given by a specific 
combination of LECs in the irreducible representation of SU(3) as summarized in 
Table 1 of \cite{XX2010} for $\La N$, $\Si N$ and for $\Xi\La$, $\Xi\Si$, $\Xi\Xi$,
and in Table 1 of \cite{YY2015} for $S=-2$. The additional SU(3) breaking terms 
($C^\chi$) that arise at NLO have been worked out and summarized in Table 10 of 
\cite{Petschauer:2013}. Those enter in various combinations into the different
BB channels and, in general, cannot be easily disentangled. Only for $^1S_0$ channels that
are pure $\{27\}$ states they can be cast into a compact form \cite{XX2015}: 
\begin{eqnarray}
\nonumber
V^{I=1}_{NN} &=& {\tilde C^{27}}+{C^{27}}(p^2+p'^2) \ +\frac{1}{2}{ C^\chi_1}(m^2_K-m^2_\pi), \\
\nonumber
V^{I=3/2}_{\Si N} &=& {\tilde C^{27}}+{C^{27}}(p^2+p'^2) \ +\frac{1}{4}{C^\chi_1}(m^2_K-m^2_\pi), \\
\nonumber
V^{I=2}_{\Si \Si} &=& {\tilde C^{27}}+{C^{27}}(p^2+p'^2), \\
\nonumber
V^{I=3/2}_{\Xi \Si} &=& {\tilde C^{27}}+{C^{27}}(p^2+p'^2) \ +\frac{1}{4}{C^\chi_2}(m^2_K-m^2_\pi), \\
V^{I=1}_{\Xi \Xi} &=& {\tilde C^{27}}+{C^{27}}(p^2+p'^2) \ +\frac{1}{2}{C^\chi_2}(m^2_K-m^2_\pi),
\label{Eq:SU32} 
\end{eqnarray}
As a consequence of that, the LEC $C^\chi_1$ can be easily fixed from a combined
study of $pp$ and $\Si^+p$ scattering \cite{XX2015}. However, there is no practical 
way to determine the altogether 6 ($^1S_0$) + 6 ($^3S_1$) SU(3) symmetry breaking 
LECs~\cite{Petschauer:2013} from the available data for the $NN$, $\La N$, $\Si N$, and 
$S=-2$ systems.  
Note that the SU(3) symmetry breaking in the meson-exchange part, 
caused by the mass differences between 
the $\pi$, $K$, and $\eta$ mesons, is taken into account in all our calculations. 

\section{Formalism for evaluating the two-particle momentum correlation function}

Details for calculating the two-particle momentum correlation function $C(k)$ within 
the Koonin-Pratt formalism can be found, e.g., in \cite{Cho:2017}, and 
for the case of coupled channels in \cite{Haidenbauer:2018}. 
The inclusion of the Coulomb interaction is discussed in \cite{Haidenbauer:2020F}. 
Below we only summarize the formulae used here. 
We assume that the correlations are primarily due to the interaction in the $S$-waves.
Thus, only the contributions from the $^1S_0$ and $^3S_1$ partial waves are
considered, though for the latter the coupling to the $^3D_1$ state is
taken into account. For an $S$-wave the correlation function is given by
\begin{equation}
C(k) = 1 + \int_0^\infty 4\pi r^2 dr\,S_{12}(r) 
\left[
\left|\psi(k,r)\right|^2
-\left|j_0(kr)\right|^2
\right] ,
\label{Eq:LL1}
\end{equation}
where $k$ is the center-of-mass momentum in the two-body system. 
$S_{12}$ is the so-called source function \cite{Cho:2017} for which 
we adopt the usual static approximation and represent the source by 
a spherically symmetric Gaussian distribution,
$S_{12}({\bf r})=\exp(-r^2/4R^2)/(2\sqrt{\pi}R)^3$, 
so that it depends only on a single parameter, the source radius $R$. 
$\psi(k,r)$ is the scattering wave function that can be obtained by 
solving the Schr\"odinger or Lippmann-Schwinger equation for a given 
potential, and $j_0(kr)$ is the spherical Bessel function for $l=0$.
When there are coupled channels one has to use the corresponding 
coupled-channel wave functions \cite{Haidenbauer:2018} 
\begin{eqnarray}
|\psi(k,r)|^2 \to \sum_\beta \omega_\beta |\psi_{\beta\alpha}(r)|^2
\label{Eq:Wave}
\end{eqnarray} 
where the sum $\beta$ runs over all two-body channels that couple to the 
state $\alpha$.
The quantity $\omega_\beta$ is their weight with 
$\omega_\alpha = 1$.
In the actual calculation we assume the spin states to contribute with 
the same weight as for free scattering, 
$C_{\rm th} (k)  = (1/4) C_{^1S_0}(k) + (3/4) C_{^3S_1}(k)$, 
and evaluate the actual correlation function via 
\begin{equation} 
C(k) = ({a} + {b}\, k) \, (1 + {\lambda} \, (C_{\rm th} (k) -1))
\label{Eq:LL2} 
\end{equation}
where $\lambda$ is the so-called impurity (or feed-down) parameter \cite{Cho:2017} 
and the polynomial factor accounts for normalization and non-femtoscopic 
background effects \cite{ALICE:2019,Kamiya:2021}. 

\section{Results}

We start with results for the BB interaction with strangeness $S=-2$ 
and specifically for the $\Xi N$ system. In this case chiral potentials 
up to NLO have been already established by us in 2016 \cite{YY2015} 
and 2019 \cite{YY2019}, respectively. 
Thereby constraints from the $\La\La$ scattering length in the $^1S_0$ state 
together with experimental upper bounds on the cross sections for $\Xi N$ 
scattering and for the transition $\Xi N \to \La\La$ have been exploited.  
This allowed us to fix the additional LECs that arise in the $\{1\}$
irreducible representation of SU(3) \cite{YY2015}. Furthermore, 
the consideration of those empirical constraints necessitated to add 
SU(3) symmetry breaking contact terms in other irreps 
($\{27\}$, $\{10\}$, $\{10^*\}$, $\{8_s\}$, $\{8_a\}$),
with regard to those determined from the $\La N$ and $\Si N$ data. 
This is anyway expected and fully in line with the power counting of 
SU(3) chiral EFT, as discussed in Sect.~2. Note that the interaction from 
2019 is more attractive in the $^3S_1$ partial wave with isospin $I=1$ \cite{YY2019}. 
Specifically, it yields a moderately attractive (in-medium) $\Xi$-nuclear interaction 
and supports the existence of bound $\Xi$-hypernuclei \cite{Le:2021X}, 
in line with experimental evidence \cite{Nakazawa:2015,Yoshimoto:2021}.
The interactions in the ($I=0,1$) $^1S_0$ partial waves are the same in the two versions.

\begin{figure}[htbp] 
 \begin{center}
   \hspace{0.3cm}{
 \includegraphics[width=0.48\textwidth,clip]{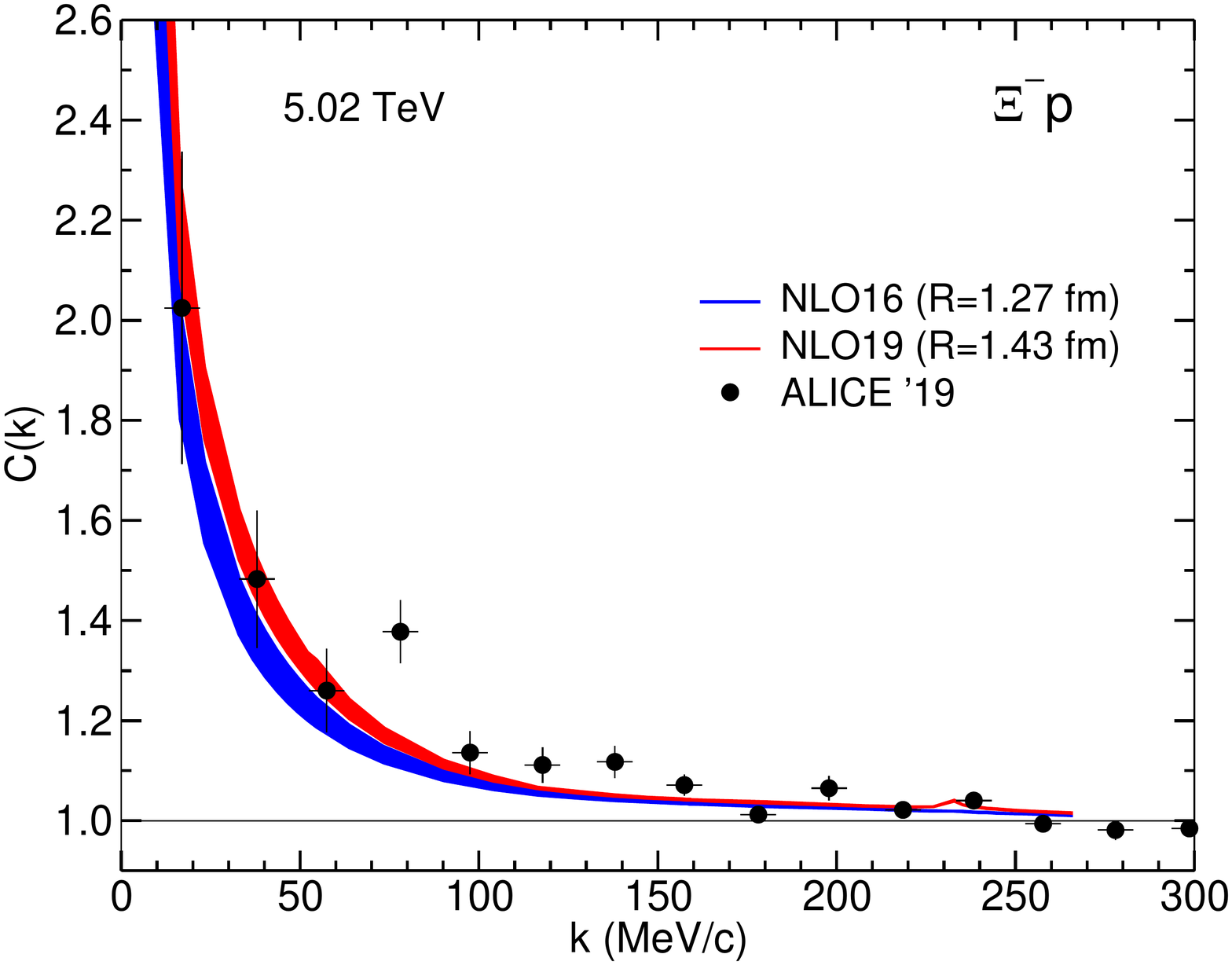}
 \includegraphics[width=0.48\textwidth,clip]{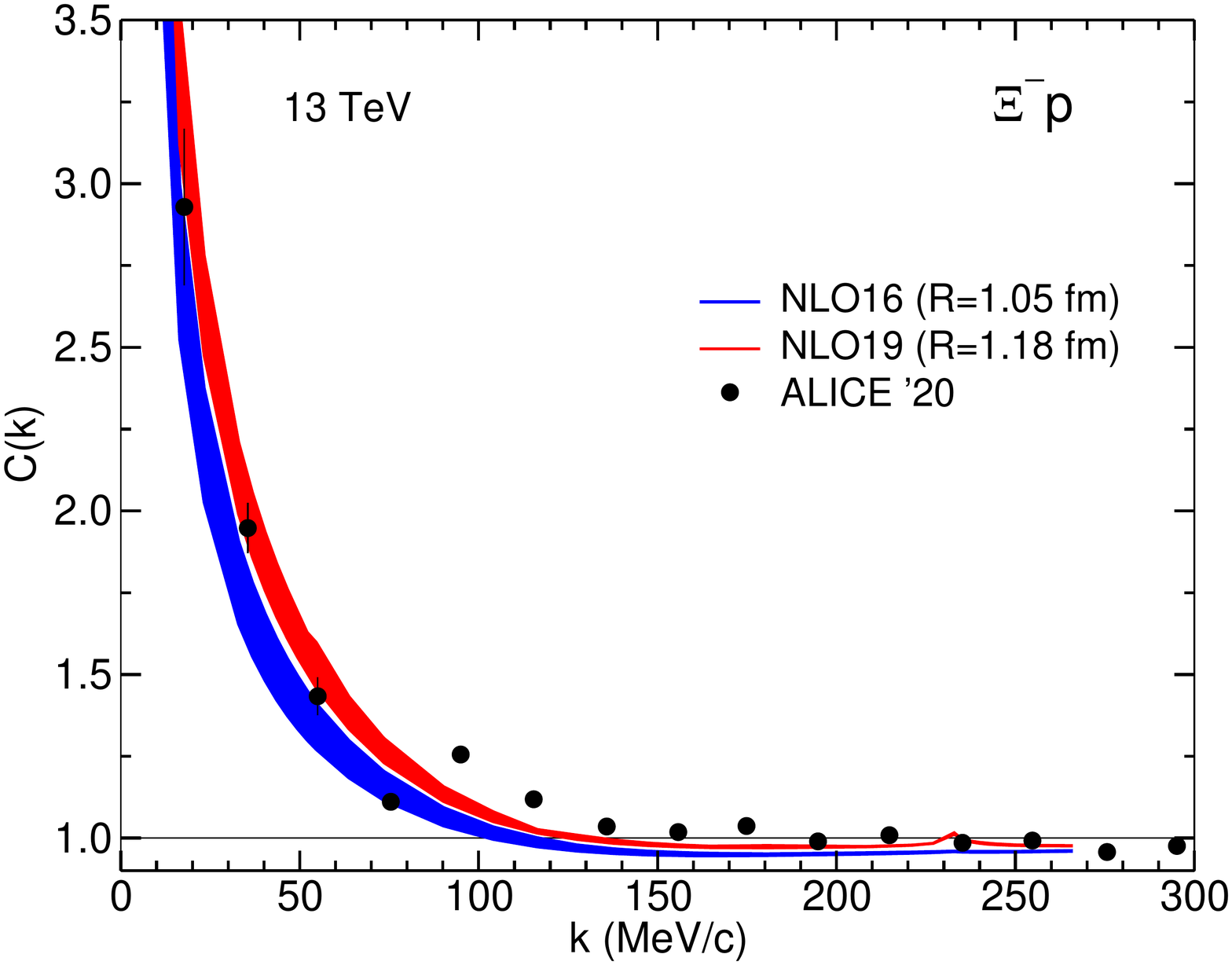}}       
 \end{center}
\caption{$\Xi^-p$ correlation functions measured in $p$-Pb collisions at 
$5.02$~TeV \cite{ALICE:2019} (left) and in $pp$ collisions at 
$13$~TeV \cite{ALICE:2020} (right). Results are shown for the NLO16 and NLO19 potentials.}
 \label{fig:XiN}
\end{figure}

Two-body momentum correlation functions for $\Xi^- p$ have been measured by the 
ALICE Collaboration in $p$-Pb collisions at $5.02$ TeV \cite{ALICE:2019}
and in $pp$ collisions at $13$ TeV \cite{ALICE:2020}. Those data are 
shown in Fig.~\ref{fig:XiN} and compared with the predictions
based on our interactions. There are also new but still preliminary 
results from Au+Au collisions at $200$ GeV by the STAR 
Collaboration \cite{STAR:2021}.   

One of the crucial ingredients in the evaluation of the correlation function
is the value of the source radius $R$, see Eq.~(\ref{Eq:LL1}). Among
other things it depends on the reaction and also on the reaction energy. In the
initial works of the ALICE Collaboration it was assumed that $R$ is the same
for all final BB systems produced in a specific collision at a 
specific energy. Accordingly, the value of $R$ was calibrated from data 
on the $pp$ correlation function measured in parallel, where the pertinent 
two-body interaction is very well established, and then this value was used for 
analyzing data on other systems like $\La N$, $\La\La$, $\Xi N$, etc. 
Recently, a model has been applied that allows them to relate the 
source radii for different final states with each other \cite{ALICE:2020R}. 
On the other hand, in the studies of Kamiya et al. \cite{Kamiya:2021,Kamiya:2019},  
$R$ is considered essentially as a free parameter that should be determined, 
case by case, directly from the correlation data in the course of the analysis. 
This strategy has been applied in their recent work on the $\La\La$ and $\Xi^-p$ 
systems \cite{Kamiya:2021} 
but, e.g., also for the $K^- p$ interaction \cite{Kamiya:2019}. In our own 
calculations we allow likewise for some flexibility in the choice of $R$. 

In Fig.~\ref{fig:XiN} we present predictions for $C(k)$ for the $S=-2$ interactions 
from 2016 (NLO16) and 2019 (NLO19), respectively. We show the results as bands
which reflect the uncertainty due to the residual cutoff dependence of the
chiral interactions \cite{YY2015,YY2019}. 
Interestingly, the correlation function calculated for NLO19
with $R$ taken from the corresponding $pp$ fits by ALICE~\cite{ALICE:2019S} 
($1.43$~fm for $5.02$~TeV and $1.18$~fm for $13$~TeV) agree nicely with 
the measurements, cf. the red bands. 
Note that in case of the $p$-Pb data a correction to the baseline has been applied, 
following Ref.~\cite{Kamiya:2021}, and the feed-down parameter $\lambda = 0.513$
has been adopted, see Eq.~(\ref{Eq:LL2}).  
For the $13$~TeV data $\lambda = 1$ is used. 
 
Results for the more weakly attractive NLO16 potential (blue bands) 
are also close to the data, provided one uses the source radii deduced 
in the work by Kamiya et al.~\cite{Kamiya:2021}
($1.27$~fm for $5.02$~TeV and $1.05$~fm for $13$~TeV).
There a $S=-2$ potential established from lattice QCD simulations
by the HAL QCD Collaboration \cite{Sasaki:2020} has been employed.
A recent study of Liu et al.~\cite{Liu:2022} uses a LO interaction based on covariant $\chi$EFT 
that was fitted to the phase shifts of the aforementioned HAL QCD potential.
In this case similar $R$ values as those adopted by us for our NLO19 results 
seem to be preferred for getting agreement with the ALICE data and not 
those of \cite{Kamiya:2021}.   

\begin{figure}[htbp] 
      \begin{center}
      \hspace{0.3cm}{
 \includegraphics[width=0.48\textwidth,clip]{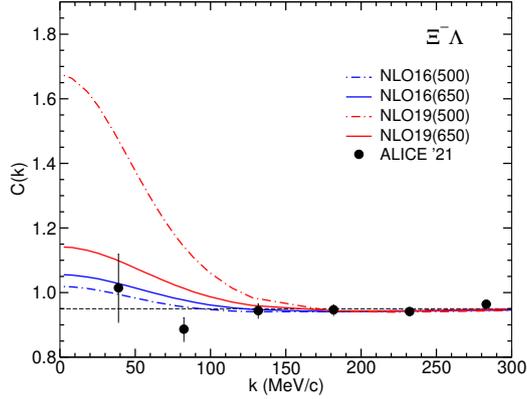}
     }       
      \end{center}
\caption{$\Xi^-\La$ correlation functions for different $\Xi\La$
interactions, see text.  
Preliminary data from $pp$ collisions at $13$~TeV are by the ALICE Collaboration \cite{ALICE:2021}.}
    \label{fig:XiL}
\end{figure}

We note that our calculation includes all relevant wave functions
of the channels that can couple to $\Xi^-p$. Thus, besides the actual $\Xi^-p$
wave function, those from the coupling to channels that are already open 
($\Xi^0 n$ and $\La\La$) are included too. Furthermore, 
components due to the coupling of $\Xi N$ to the $\La\Si$ channel 
(which opens at around $k\approx 230$~MeV/c) are taken into account. As a result of 
that there is a small but visible cusp in the predicted correlation function 
at the $\Xi N$ threshold, see Fig.~\ref{fig:XiL}. 
(See also Ref.~\cite{Liu:2022} in this context.)
There is no visible cusp for NLO16 because here the ($I=1$) $^3S_1$ interaction 
is weaker/repulsive \cite{YY2019}. 
A cusp appears also in 
the $\La p$ correlation function at the opening of the $\Si N$ threshold and has
been experimentally established in a recent measurement by the ALICE Collaboration
\cite{ALICE:2021L}. 
Whether the much less pronounced cusp predicted for $\Xi^-p$ can be 
experimentally observed/confirmed remains to be seen. 
In the present calculation all components are included with the weight $\omega_\beta=1$ 
and by assuming the same source function as for the diagonal channel~\cite{Haidenbauer:2018}.  


Recently the ALICE Collaboration presented first results for the 
correlation function of the $S=-3$ system $\Xi^-\La$ \cite{ALICE:2021}. 
The data, obtained in $pp$ collisions at $13$~TeV, are still very preliminary. 
Thus, only qualitative conclusions can be drawn at the present stage. 
Nevertheless, the fact that $C(k)$ is close to the baseline (dashed curve) over the 
whole momentum region, see Fig.~\ref{fig:XiL}, strongly suggests that 
the interaction in the $\Xi^-\La$ system should be weak, though a scenario 
like that observed for $\Si^+ p$  \cite{Haidenbauer:2021} cannot be excluded at
this stage. Since 
$C(k)$ at the lowest momentum is well above the baseline one 
is tempted to conclude that the $\Xi\La$ interaction is overall weakly 
attractive \cite{Cho:2017}. 
Some phenomenological BB potentials in the literature predict 
such a weak $\Xi\La$ interaction (NSC97a \cite{Stoks:1999}, 
fss2 \cite{Fujiwara:2006}). Studies within $\chi$EFT, so far 
performed only at LO, lead, however, to strongly attractive
forces in the $S=-3$ sector, and even support the existence of 
bound states in some spin-isospin channels \cite{XX2010,Liu:2020}. 
The preliminary ALICE data practically rule out any $\Xi\La$ bound 
states and make clear that the simple extension of LO potentials, fitted 
to $\La N$ and $\Si N$ data, to $S=-3,-4$ based on strict SU(3) 
symmetry is certainly unrealistic.

In Fig.~\ref{fig:XiL} we present results for the $\Xi^-\La$ correlation function
based on an NLO $S=-3$ interaction with LECs fixed in the $S=-2$ sector. 
This choice accounts for the overall trend that the BB interaction 
becomes gradually less attractive with increasing $|S|$, but ignores a 
possible SU(3) symmetry breaking in the contact terms between $S=-2$ and $S=-3$.   
Nonetheless, it is instructive to see the predictions of such an interaction.
The results are for the cutoffs $\La = 500$ and  $650$~MeV \cite{YY2019}. 
We use the parameters given by the ALICE Collaboration, namely $R=1.03$~fm
and $\lambda=0.36$ \cite{ALICE:2021}, and we multiply our results with 
$a\approx 0.95$,  
cf. Eq.~(\ref{Eq:LL2}), to correct for the shifted baseline. 
The blue (solid and dash-dotted) curves are results with LECs taken over from the NLO16
interaction~\cite{YY2015}. As discussed above, the NLO16 $\Xi N$ interaction 
is possibly too weak. However, the predicted $\Xi^-\La$ correlation functions
are perfectly in line with the preliminary ALICE data, see Fig.~\ref{fig:XiL}. 
The $\Xi\La$ scattering lengths are $a_{s} = -0.99 \ccc -0.89$ fm ($^1S_0$)
and $a_{t} = 0.026 \ccc -0.12$ fm ($3S_1$). 
For LECs taken from the more attractive NLO19 potential, the $\Xi^-\La$ correlation 
function is larger and, moreover, there is also a sizable cutoff dependence, cf. 
the red (solid and dash-dotted) lines. There is also a small cusp at the opening of
the $\Xi\Si$ channel (at around $k\approx 310$~MeV/c). 
The $^3S_1$ scattering lengths are $a_{t} = -0.42 \ccc -1.66$ fm.
Those for the $^1S_0$ are the same as for NLO16. 
Anyway, once the analysis of the experiment is finalized, quantitative and reliable
conclusions on the actual strength of the $\Xi^-\La$ interaction are possible. 
Then one can try to fix the LECs for the SU(3) symmetry breaking contact terms,
appropriate for the extension to $S=-3$.
 
Finally, we consider the $\Xi^-\Xi^-$ system where experiments have been performed
by the STAR Collaboration. Also those data, obtained in Au+Au collisions at 
$200$~GeV \cite{STAR:2021}, are still preliminary. 
For systems with two identical particles the correlation function involves 
an additional term from quantum statistics \cite{Cho:2017,Haidenbauer:2018},  
besides the contribution from the actual interaction,  
which in case of two fermions yields a suppression of the correlation at 
low momenta due to the anti-symmetrization of the wave function. 
The presence of a repulsive Coulomb interaction in that
system leads to an even stronger suppression. Nonetheless for a strongly attractive
hadronic interaction as in the $^1S_0$ partial wave of the $pp$ system the 
corresponding correlation function shows a pronounced peak at moderate momenta with values
well above the baseline of $C(k)=1$ \cite{ALICE:2019S}. The preliminary data reported 
by STAR suggest that $C(k) \le 1$ in case of $\Xi^-\Xi^-$ for all momenta, which 
indicates that the $\Xi\Xi$ interaction in the $^1S_0$ state could be much less
attractive than that in $pp$.  

\begin{figure}[htbp] 
      \begin{center}
      \hspace{0.3cm}{
 \includegraphics[width=0.48\textwidth,clip]{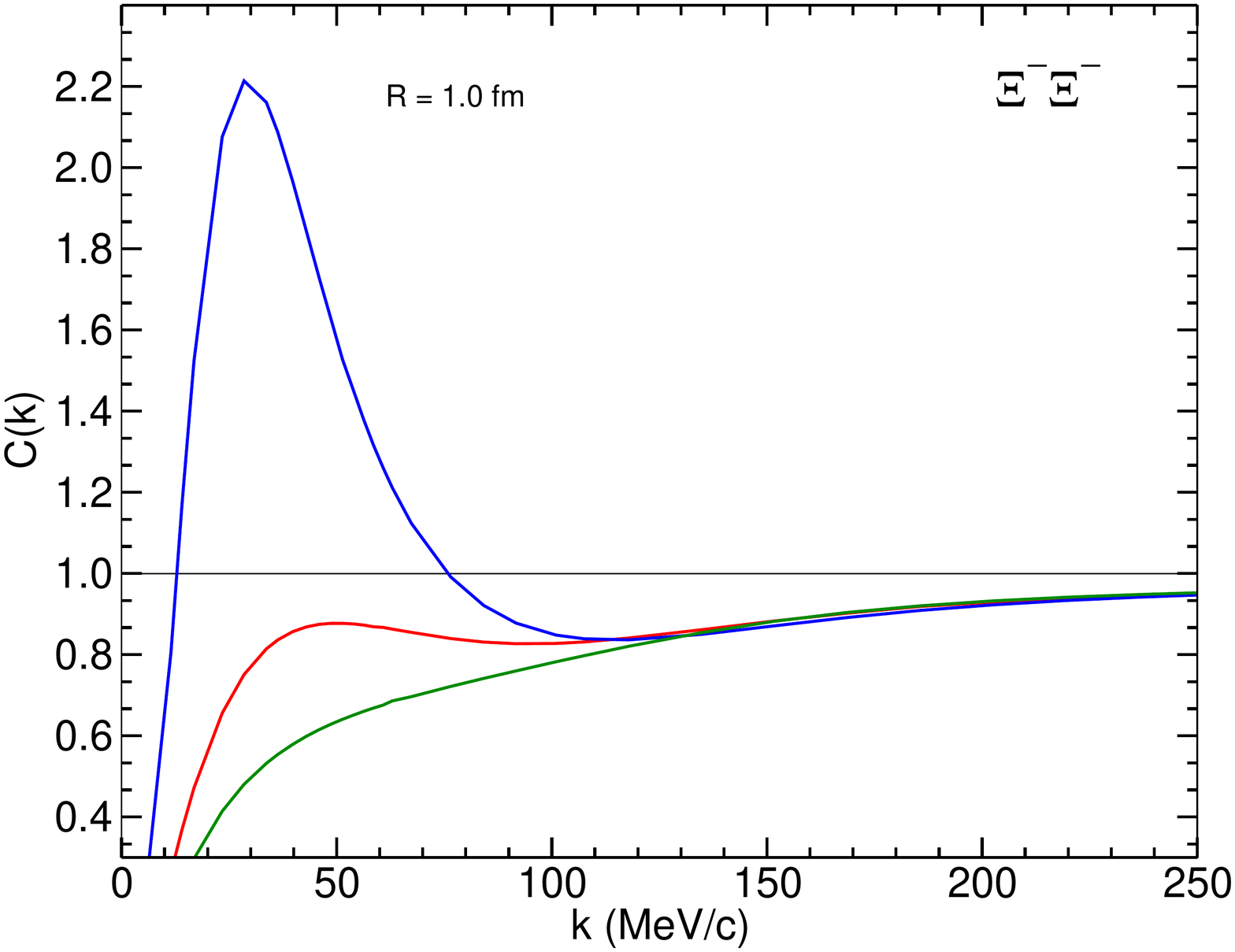}
 \includegraphics[width=0.48\textwidth,clip]{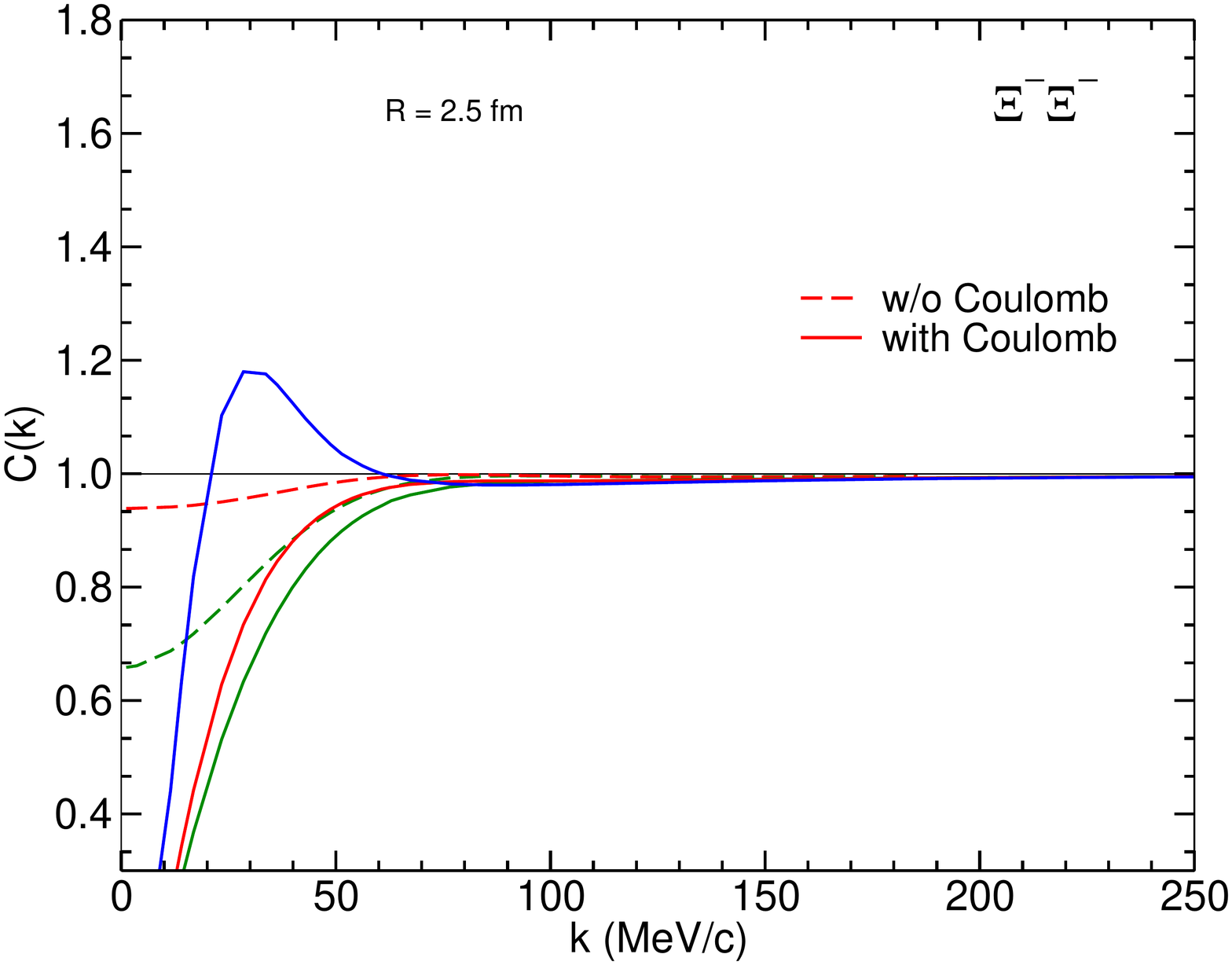}}       
      \end{center}
\caption{$\Xi^-\Xi^-$ correlation functions for different $\Xi\Xi$
interactions, see text. The results are for a source radius of $R=1.0$~fm (left) 
and of $2.5$~fm (right), respectively.}
    \label{fig:XiXi}
\end{figure}

Our results for the $\Xi^-\Xi^-$ correlation function are presented in 
Fig.~\ref{fig:XiXi}. As mentioned, contrary to the systems discussed above, 
the SU(3) structure and also SU(3) symmetry breaking is very simple for this 
channel because it is a pure $\{27\}$ state and, thus, closely related to the 
$^1S_0$ interactions in the $pp$, $\Si^+p$, and $\Si^+\Si^+$ systems. Indeed, 
if SU(3) symmetry would be strictly fulfilled, all those interactions are identical, 
see Eq.~(\ref{Eq:SU32}). This feature has been exploited in
Ref.~\cite{XX2015} to estimate the pertinent SU(3) symmetry breaking contact terms
and to make predictions for the corresponding $\Xi\Xi$ interaction. Nevertheless,
the decisive LEC $C^\chi_2$ could not be pinned down quantitatively and three different
scenarios for its magnitude were considered in that work. We show here predictions
for those scenarios, for the cutoff $\La = 500$~MeV \cite{XX2015}.    

The source radii appropriate for correlation functions measured in heavy ion collisions 
as done by the STAR Collaboration are typically in the order of $R=2.5 - 5$~fm. 
In our illustrative calculation we use $R=2.5$~fm and $\lambda=1$~fm (right). 
For comparison and in order to stimulate measurements in $pp$ collisions we present 
also predictions based on a source radius of $R= 1$~fm (left). 
The blue curves are results with the $\{27\}$ LECs deduced for $\Si^+\Si^+$ in \cite{XX2015}.  
It corresponds to the assumption that there is no further SU(3) symmetry breaking in 
the contact terms beyond $S=-2$. In this case the strength of the $\Xi\Xi$ interaction 
is comparable to that for $pp$, see the phase shifts in Fig.~4 in Ref.~\cite{XX2015}, 
and the shape of the correlation function (Fig.~\ref{fig:XiXi}) is also similar to that 
for $pp$ \cite{ALICE:2019S}. 
Assuming that the symmetry breaking is as large as that required for a consistent 
description of $pp$ and $\Si^+p$ leads to the green curves. The may be most realistic 
assumption that the breaking is roughly half way between yields the red curves. 
For the latter two scenarios the correlation 
function remains below the baseline of $C(k)=1$. Indeed, such a behavior is supported
by the preliminary STAR data. The dashed lines indicate results without the Coulomb
interaction for the latter two scenarios. One can see that already in this case 
$C(k) < 1$ because of the quantum statistical effect.    
Note that lattice QCD simulations from the HAL QCD Collaboration, for almost physical 
masses ($m_\pi \approx 146$~MeV), suggest phase shifts of the $^1S_0$ 
state within the range spanned by the latter two scenarios \cite{Doi:2018}.  
   
\section{Summary}

The J\"ulich-Bonn-Munich Collaboration has applied chiral effective field theory 
to investigate the baryon-baryon interaction involving hyperons.
These studies, performed so far up to next-to-leading order (NLO) in the chiral
expansion, have shown that for the strangeness $S=-1$ ($\Lambda N$, $\Sigma N$) [1,2]
and $S=-2$ ($\Lambda \Lambda$, $\Xi N$) [3,4] sectors a consistent and satisfactory description 
of the available scattering data and experimental constraints can be achieved within the 
assumption of broken SU(3) flavor symmetry. 

In the present contribution we have discussed a possible extension of this approach to 
strangeness $S=-3$ and $S=-4$ baryon-baryon systems where empirical information is 
rather scarce. Specifically we have shown that measurements of two-particle
correlation functions in heavy-ion collisions and/or in high energetic 
proton-proton collisions can be used to constrain the interaction in channels 
like $\Xi\Lambda$ or $\Xi\Xi$. Pertinent measurements are already on their way 
\cite{ALICE:2021,STAR:2021} and, hopefully, experimental studies of further
$S=-3$ and $-4$ systems will follow.  

\vskip 0.4cm \noindent
{\bf Acknowledgements}:
This work is supported in part by the DFG and the NSFC through
funds provided to the Sino-German CRC 110 ``Symmetries and
the Emergence of Structure in QCD'' (DFG grant. no. TRR~110)
and the VolkswagenStiftung (grant no. 93562).
The work of UGM was supported in part by the Chinese Academy
of Sciences (CAS) President's International Fellowship Initiative (PIFI)
(grant no.~2018DM0034) and by the European Research Council (ERC) under the
European Union's Horizon 2020 research and innovation programme (ERC AdG EXOTIC,
grant agreement No. 101018170).


\begin{thebibliography}{99}

\bibitem{Epelbaum:2009}
  E.~Epelbaum, H.-W.~Hammer, U.-G.~Mei{\ss}ner,
  Rev. Mod.\ Phys.\ {\bf 81}, 1773 (2009).

\bibitem{Reinert:2017} 
  P.~Reinert, H.~Krebs and E.~Epelbaum,
  Eur.\ Phys.\ J.\ A {\bf 54}, 86 (2018).

\bibitem{Entem:2017}
D.~R.~Entem, R.~Machleidt and Y.~Nosyk,
Phys. Rev. C \textbf{96}, 024004 (2017).


\bibitem{YN2013}
J. Haidenbauer, S. Petschauer, N. Kaiser, U.-G. Mei\ss ner, A. Nogga, and W. Weise,
Nucl. Phys. A {\bf 915}, 24 (2013). 

\bibitem{YN2019}
J.~Haidenbauer, U.-G.~Mei\ss{}ner, A.~Nogga,
Eur. Phys. J. A \textbf{56}, 91 (2020). 

\bibitem{YY2015}
J.~Haidenbauer, U.-G.~Mei\ss{}ner, S.~Petschauer,
Nucl. Phys. A \textbf{954}, 273 (2016).
 
\bibitem{YY2019}
J.~Haidenbauer and U.-G.~Mei\ss{}ner, Eur. Phys. J. A \textbf{55}, 23 (2019). 


\bibitem{Le:2020}
H.~Le, J.~Haidenbauer, U.-G.~Mei\ss{}ner and A.~Nogga,
Eur. Phys. J. A \textbf{56}, 301 (2020).

\bibitem{Le:2021LL}
H.~Le, J.~Haidenbauer, U.-G.~Mei\ss{}ner and A.~Nogga,
Eur. Phys. J. A \textbf{57}, 217 (2021).


\bibitem{Ishii:2018}
N.~Ishii \textit{et al.} [HAL QCD], 
EPJ Web Conf. \textbf{175}, 05013 (2018).

\bibitem{Doi:2018}
T.~Doi \textit{et al.} [HAL QCD],
EPJ Web Conf. \textbf{175}, 05009 (2018).

\bibitem{NPLQCD:2021}
M.~Illa \textit{et al.} [NPLQCD],
Phys. Rev. D \textbf{103}, 054508 (2021).



\bibitem{XX2010}
J.~Haidenbauer and U.-G.~Mei\ss{}ner,
Phys. Lett. B \textbf{684}, 275 (2010).

\bibitem{Liu:2020}
Z.-W.~Liu, J.~Song, K.-W.~Li and L.-S.~Geng,
Phys. Rev. C \textbf{103}, 025201 (2021).

\bibitem{Petschauer:2013}
S.~Petschauer and N.~Kaiser,
Nucl. Phys. A \textbf{916}, 1 (2013).

\bibitem{XX2015}
J.~Haidenbauer, U.-G.~Mei\ss{}ner and S.~Petschauer,
Eur. Phys. J. A \textbf{51}, 17 (2015). 

 

\bibitem{Cho:2017} 
  S.~Cho {\it et al.} [ExHIC],
  Prog.\ Part.\ Nucl.\ Phys.\  {\bf 95}, 279 (2017). 

 \bibitem{Adams:2006}
  J.~Adams {\it et~al.},
  Phys. Rev. C {\bf 74}, 064906 (2006).

\bibitem{Adamczewski:2016}
  J.~Adamczewski-Musch {\it et al.} [HADES],
  Phys.\ Rev.\ C {\bf 94}, 025201 (2016).

\bibitem{ALICE:2021L}
S.~Acharya \textit{et al.} [ALICE],
[arXiv:2104.04427 [nucl-ex]].

\bibitem{STAR:2015}
  L.~Adamczyk {\it et al.} [STAR],
  Phys.\ Rev.\ Lett.\  {\bf 114}, 022301 (2015).

\bibitem{ALICE:2019S}
S.~Acharya \textit{et al.} [ALICE],
Phys. Lett. B \textbf{797}, 134822 (2019).

\bibitem{ALICE:2019}
S.~Acharya \textit{et al.} [ALICE],
Phys. Rev. Lett. \textbf{123}, 112002 (2019).

\bibitem{ALICE:2020}
S.~Acharya \textit{et al.} [ALICE],
Nature \textbf{588}, 232 (2020).

\bibitem{ALICE:2021} E.~S. Chizzali [ALICE],
Conference on Strangeness in Quark Matter Conference 2021 (SQM2021),
{\tt
https://indico.cern.ch/event/985652/contributions/4305105/
}

\bibitem{STAR:2021} M. Isshiki [STAR],
[arXiv:2109.10953 [nucl-ex]].


\bibitem{Polinder:2006}
H.~Polinder, J.~Haidenbauer and U.-G.~Mei{\ss}ner,
Nucl. Phys. A \textbf{779}, 244 (2006).

\bibitem{Polinder:2007}
H.~Polinder, J.~Haidenbauer and U.-G.~Mei{\ss}ner,
Phys. Lett. B \textbf{653}, 29 (2007).

\bibitem{Petschauer:2020}
S.~Petschauer, J.~Haidenbauer, N.~Kaiser, U.-G.~Mei\ss{}ner and W.~Weise,
Front. in Phys. \textbf{8}, 12 (2020).

\bibitem{Haidenbauer:2018}
J.~Haidenbauer,
Nucl. Phys. A \textbf{981}, 1 (2019).

\bibitem{Haidenbauer:2020F}
J.~Haidenbauer, G.~Krein and T.~C.~Peixoto,
Eur. Phys. J. A \textbf{56}, 184 (2020).

\bibitem{Kamiya:2021}
Y.~Kamiya, K.~Sasaki, T.~Fukui, T.~Hyodo, K.~Morita, K.~Ogata, A.~Ohnishi and T.~Hatsuda,
[arXiv:2108.09644 [hep-ph]].

\bibitem{Le:2021X}
H.~Le, J.~Haidenbauer, U.-G.~Mei\ss{}ner and A.~Nogga,
Eur. Phys. J. A \textbf{57}, 339 (2021).

\bibitem{Nakazawa:2015}
K.~Nakazawa \textit{et al.}, 
PTEP \textbf{2015}, 033D02 (2015).

\bibitem{Yoshimoto:2021}
M.~Yoshimoto \textit{et al.},
PTEP \textbf{2021}, 7 (2021).

\bibitem{ALICE:2020R}
S.~Acharya \textit{et al.} [ALICE],
Phys. Lett. B \textbf{811}, 135849 (2020). 


\bibitem{Kamiya:2019}
Y.~Kamiya, T.~Hyodo, K.~Morita, A.~Ohnishi and W.~Weise,
Phys. Rev. Lett. \textbf{124}, 132501 (2020).


\bibitem{Sasaki:2020}
K.~Sasaki \textit{et al.} [HAL QCD],
Nucl. Phys. A \textbf{998}, 121737 (2020).


\bibitem{Liu:2022}
Z.-W.~Liu, K.-W.~Li and L.-S.~Geng,
[arXiv:2201.04997 [hep-ph]].

\bibitem{Haidenbauer:2021}
J.~Haidenbauer and U.-G.~Mei\ss{}ner,
[arXiv:2109.11794 [nucl-th]].

\bibitem{Stoks:1999}
  V.~G.~J.~Stoks and T.~A.~Rijken,
  Phys.\ Rev.\  C {\bf 59}, 3009 (1999).

\bibitem{Fujiwara:2006}
  Y.~Fujiwara, Y.~Suzuki and C.~Nakamoto,
  Prog.\ Part.\ Nucl.\ Phys.\  {\bf 58}, 439 (2007).

\end{thebibliography}
\end{document}